\documentclass[fleqn,twoside]{article}
\usepackage{espcrc2}
\usepackage{epsfig}

\usepackage{graphicx}
\title{Dependence of Dirac eigenmodes on boundary conditions
for SU(2) lattice gauge theory
\thanks{Talk presented at QCD04 by C.G.
This work was supported by DFG and BMBF.}}

\author{{Christof Gattringer and
Stefan Solbrig}
\vskip2mm
Institut f{\"u}r Theoretische Physik, Universit{\"a}t
Regensburg, 93040 Regensburg, Germany.}
      
\begin{document}
\begin{abstract}
We analyze zero modes of the Dirac operator for SU(2) lattice gauge
theory. We find that the zero modes are strongly localized in all 4 
directions. The position of these lumps depends on the boundary conditions
we use for the Dirac operator. 
We compare periodic boundary conditions and anti-periodic boundary 
conditions and find that the position of the zero modes jumps for about 
one third of the configurations.   
\vspace{-2mm}
\end{abstract}
\maketitle
\section{Introduction}

In the last few years the study of Dirac eigenmodes was established as 
a powerful method for analyzing infrared structures of lattice gauge field
configurations. 
After early work (e.g.\ \cite{early}), the idea of probing
the QCD vacuum with Dirac eigenmodes picked up steam with the discussion of
the local chirality of near-zero modes \cite{locchir}. In \cite{dual}
evidence was given, that the infrared structures which couple to the 
low lying Dirac eigenmodes are self-dual, and in \cite{lowdim} 
it was argued, that these structures could
be low-dimensional objects.

A new technique, the probing of the QCD vacuum with eigenmodes that 
are computed for different boundary conditions of the Dirac operator, was 
introduced for the study of so-called Kraan van Baal (KvB) 
solutions \cite{kvb}. 
KvB solutions are solutions of the classical Yang-Mills equations
on a Euclidean cylinder. A KvB solution of topological charge $Q = 1$
can be seen to consist of $N$ constituents (for SU($N$)), which locally 
resemble monopoles. 
For higher charges one finds $Q \times N$ constituents
\cite{highercharge}. 
The connection to the Dirac eigenmodes was given in 
\cite{kvbzeromodes}, where the zero mode $\phi$ for a 
charge 1 KvB solution was 
calculated. This eigenvalue problem was solved for a general boundary
condition in the compactified (time-) direction,
\begin{equation}
\phi(x,\beta) \; = \; e^{i 2\pi \zeta} \phi(x,0) \; ,
\label{bc}
\end{equation}
where $\beta$ is the extent of the compact direction of the Euclidean cylinder.
The boundary phase $\zeta$ has values between 0 and 1 (we stress that in 
all of our discussion the gauge fields obey periodic boundary conditions). 
It was shown, that 
the zero mode is not distributed over all of the constituents simultaneously,
but instead is localized on only one of them. 
Which of the constituents is chosen depends on the 
phase $\zeta$, relative to the phases determining the Polyakov loop 
at spatial infinity. Thus if one varies $\zeta$, the zero mode will become 
located on each of the constituents, one after the other. 
This property of the zero mode was confirmed in lattice studies 
\cite{kvblat} and strong evidence for the appearance of
KvB solutions in thermalized lattice gauge configurations was given. 
In \cite{instantonoverlap} it was argued, that an alternative 
interpretation where
the zero mode hops between integer charged objects is unlikely. 
  
Subsequently the same type of analysis, i.e.\ using the boundary condition 
(\ref{bc}) to probe gauge configurations, was applied to SU(3) lattice 
gauge theory on the periodic torus \cite{torusjump}. 
Similar to the case of the Euclidean cylinder,
it was observed, that the zero mode of a charge 1 configuration
can be located at different positions, depending on the boundary condition.
The observed zero modes are localized in all four directions.  
However, for the torus no analytic solutions are available, and the
nature of the constituents is not known. The central question is whether 
also on the torus charge 1 configurations can have constituents. 
Attempts \cite{toruscool}
to create localized 
charge 1 configurations with separated constituents on the torus 
with cooling have failed so far. However, constituent-like structures
in the cooled configurations are suggested by the behavior of the 
Polyakov loop observed in \cite{toruscool}. Why these potential constituents 
are not separated in the cooled configurations is not clear. 

In \cite{margarita} zero modes for constant curvature  
configurations on the periodic torus were studied analytically. For SU(3) 
it was found that constant curvature configurations with charge 1
can be constructed such that the zero mode hops when the boundary 
condition is changed. 
Although this zero mode is not very strongly localized, constant curvature 
configurations cannot be a priori 
ruled out as candidates for explaining the findings
of \cite{torusjump}. For SU(2) the situation is different.  
It was shown that constant curvature SU(2) configurations only have 
even charge \cite{margarita}. Hence for SU(2), Ref.\ 
\cite{margarita} rules out 
constant curvature configurations as candidates 
for explaining the hopping of the zero mode for charge 1 configurations.
Thus it is interesting to know whether SU(2) configurations 
with charge 1 show zero modes that change their position as one changes 
the boundary condition. This question is addressed in this paper. 

\section{Technicalities}

Our SU(2) gauge field configurations were generated with the 
L\"uscher-Weisz action \cite{luweact} on two different lattice sizes:
$16^4$ with $\beta = 1.95$ and $12^4$ with $\beta = 1.90$.
For both volumes we generated a large number of well
decorrelated gauge field configurations. For the complete ensembles we
calculated the 30 (50 for size $12^4$) 
smallest eigenvalues and the corresponding eigenvectors
of the chirally improved lattice Dirac operator \cite{dci} with 
periodic boundary conditions. The chirally improved Dirac operator has 
good chiral properties and was shown \cite{instantontest}
to optimally resolve the zero mode of a discretized instanton. 
We use the implicitly restarted Arnoldi method
for calculating the eigenmodes.

From all configurations we subsequently 
selected the subset of configurations with a single zero mode. 
Thus, according to the index theorem,
only configurations with topological charge $Q = \pm 1$ were chosen. 
Restricting to configurations with only a single zero mode, 
allowed us to avoid potential mixing of zero modes. Such a mixing could 
occur in configurations with higher charge, since a linear combination 
of different zero modes is again a zero mode. For such a mixture the 
hopping which we observe could be a
hopping between two zero modes corresponding to integer charged
objects. The restriction to 
configurations with a single zero mode excludes this simplest mixing 
mechanism, as opposed to mixing with near-zero modes.

For the selected subset we calculate the eigenmodes and eigenvectors a 
second time, now using boundary conditions that are anti-periodic in  
the 4-direction and periodic for the other three directions
(we will refer to these mixed boundary conditions as
``anti-periodic''). For most configurations in the subset we again 
find a single zero mode also for anti-periodic boundary conditions. 
There are, however, a few exceptions which we discuss for the $16^4$ 
ensemble: In this ensemble we have 105 configurations with a single zero mode 
for periodic boundary conditions. When switching to anti-periodic boundary
conditions, for 4 of the 105 configurations we found three zero modes.
In all the 4 cases one of the three 
modes had opposite chirality, such that the topological charge remained 
unchanged. Another configuration had four zero modes,
such that the topological charge 
was different. These 5 configurations were omitted in the final analysis.
This leaves us with 100 configurations on $16^4$ where we find exactly
one zero mode for both boundary conditions. For $12^4$ we have 114 such 
configurations. These subsets were used for our analysis.

\section{Localization of the zero modes}

Let us begin the presentation of our results with an analysis of the 
scalar density of the zero modes. For a zero mode $\phi(x)_{\alpha,c}$,
the scalar density $\rho(x)$ is defined as
\begin{equation}
\rho(x) \; = \; \sum_{\alpha=1}^4 \sum_{c=1}^2 \, 
| \phi(x)_{\alpha,c} |^2  \; .
\end{equation}
Here $\alpha$ and $c$ denote the Dirac and color indices. The
scalar density is a gauge invariant object. 

In order to demonstrate that the zero modes are localized in all four 
directions, in Figs.\ 1 and 2 we show $\rho(x)$ for two different slices  
through the position of the maximum of the lump. In Fig.\ 1
this is done for the zero mode with periodic boundary conditions, 
in Fig.\ 2 for the anti-periodic boundary condition. Both figures are for the 
same configuration, but the positions of the zero modes with periodic and 
anti-periodic boundary conditions differ. For our example these positions are 
$x_P =  (10,7,7,13) \; , \; x_A = (10,1,1,7)$.
Thus the lumps seen by the periodic and the anti-periodic
zero mode are 10.39 lattice spacings apart.  

\begin{figure}[t]
\vspace*{-10mm}
\epsfig{file=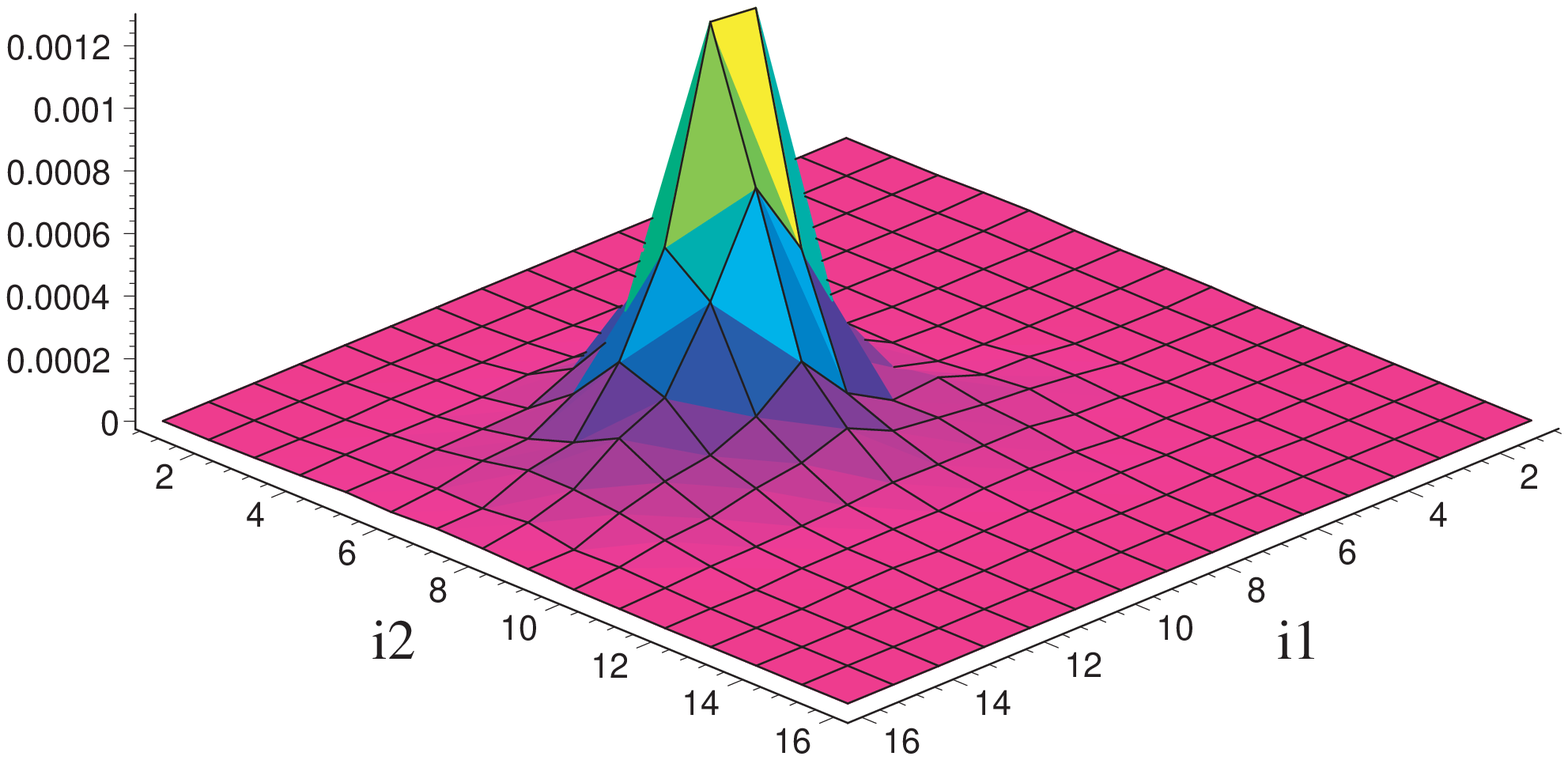,width=6.5cm,clip}
\vskip-6mm
\hspace*{10mm}
\epsfig{file=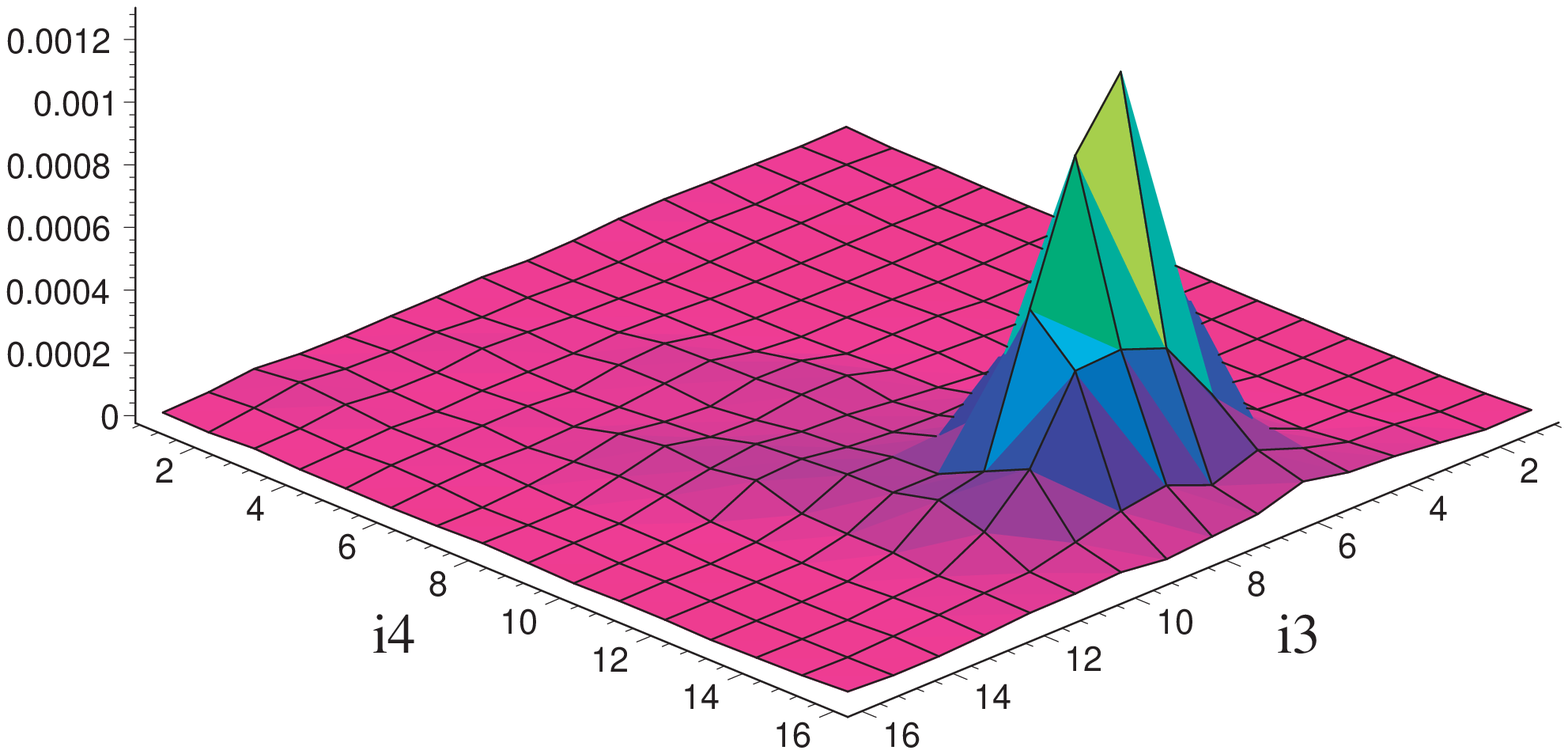,width=6.5cm,clip}
\vskip-7mm
\caption{The pseudoscalar density for a zero mode with periodic
boundary conditions. We show the 1-2 and 3-4 slices through the 
maximum of $\rho(x)$ which is
located at $(10,7,7,13)$.
\label{ph00plot}}
\vspace{-3mm}
\end{figure}

\begin{figure}[t]
\vspace*{-8mm}
\epsfig{file=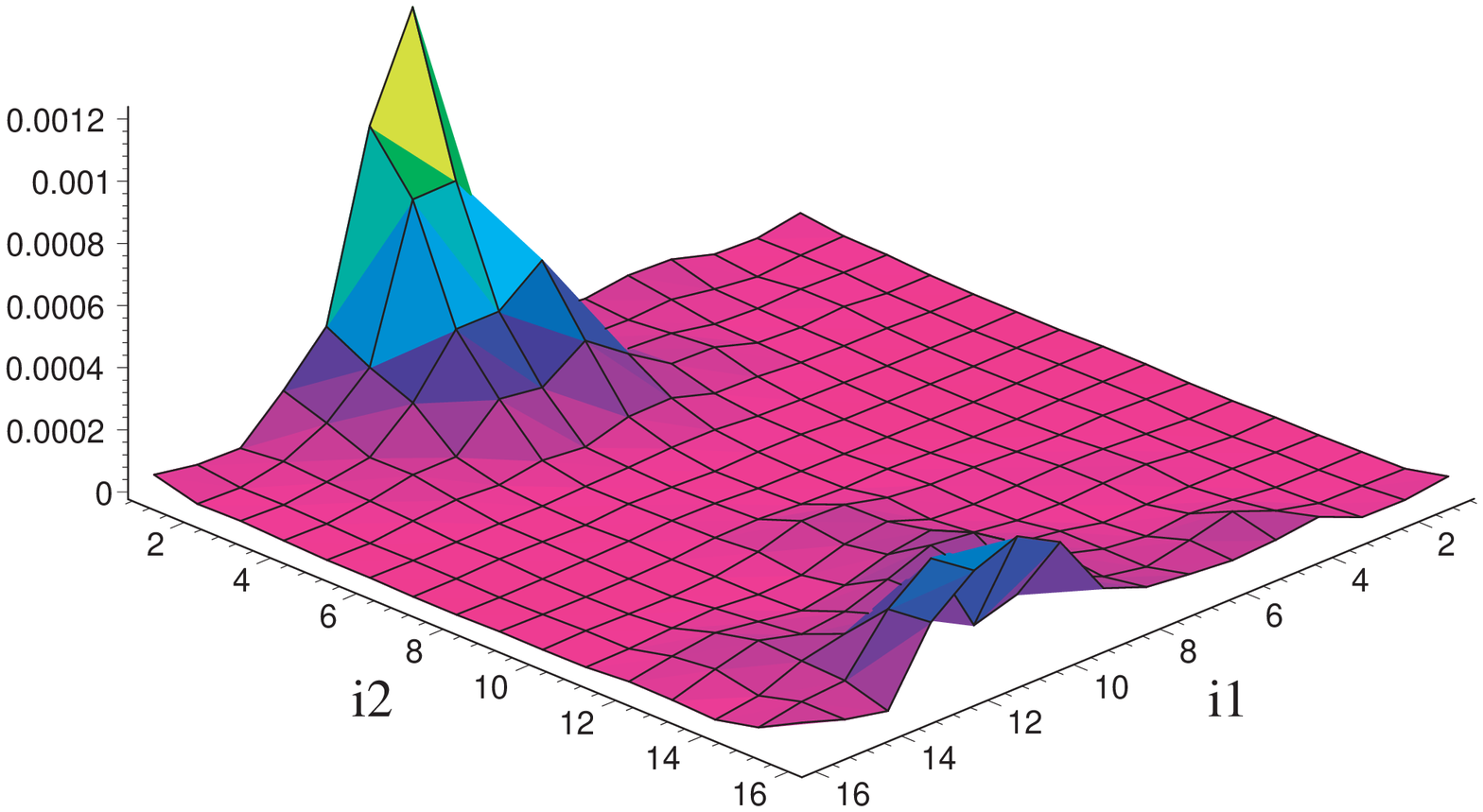,width=6.5cm,clip}
\vskip-6mm
\hspace*{10mm}
\epsfig{file=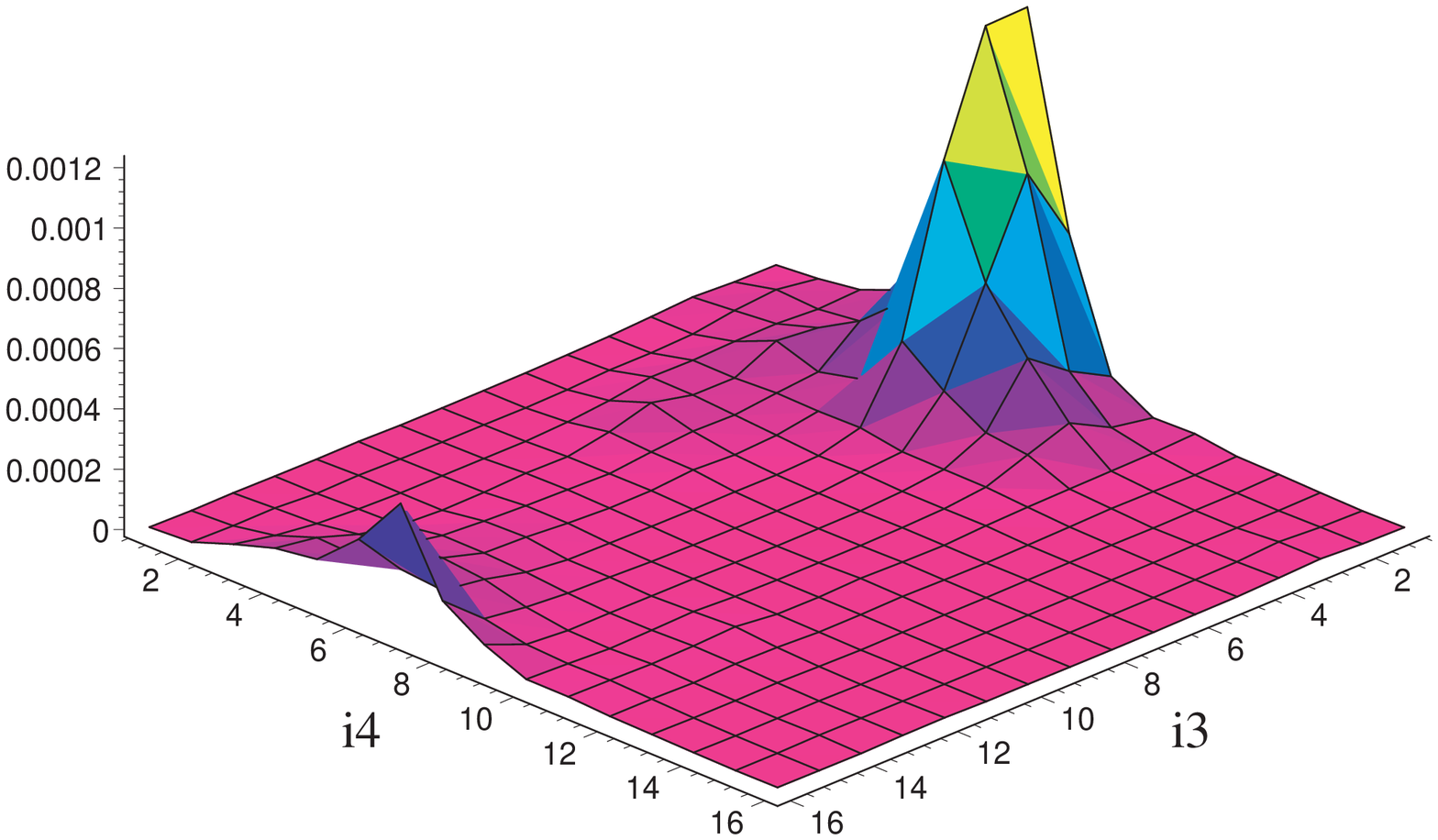,width=6.5cm,clip}
\vskip-8.8mm
\caption{Same as Fig.\ 1, but now for the zero mode
with anti-periodic boundary conditions. For 
these boundary conditions the
maximum is located at $(10,1,1,7)$.
\label{ph05plot}}
\vspace{-3mm}
\end{figure}

From the figures it is obvious, that both the periodic and the 
anti-periodic zero mode are localized in all directions. In particular we 
do not observe any special role of the 4-direction where we attached the 
anti-periodic boundary condition. 

\section{Distribution of the lump separation}

In the last section we have discussed that some of the configurations 
have the periodic and the anti-periodic zero modes at different positions. 
In order to study this phenomenon in more detail, we analyze the distance 
$d$ between the maxima of the two zero modes. We define
\begin{equation}
d \, = \, ||\, x_P - x_A \, ||_{\, torus} \; ,
\end{equation}
where $x_P, x_A$ are the positions of the maxima of $\rho(x)$ for the
zero mode with periodic and anti-periodic boundary conditions. 
The distance $||\, ... \,||_{\,torus}$ is 
the minimal Euclidean distance on the torus, i.e.\
for each component of the vector we use the shorter of the two possible
distances. 

\begin{figure}[t]
\vspace*{-2mm}
\hspace*{6mm}
\epsfig{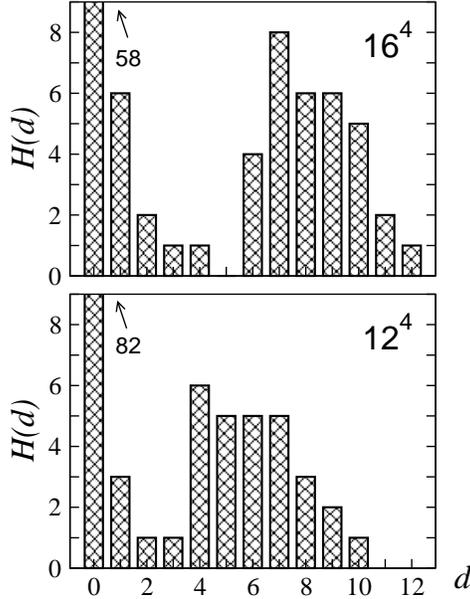}
\vskip-9mm
\caption{Histogram for the distribution of $d$.
\label{histfig}}
\vspace{-6mm}
\end{figure}

In Fig.\ 3 we show histograms for the distribution of $d$ for
the two lattice sizes. We find that for both lattice sizes a large portion
of the configurations have their periodic and anti-periodic zero modes
at the same position, or only a small distance apart. However, both
histograms show a second bump in the distribution at distances between $d = 4$
and $d = 12$. In particular we find that for the larger lattice 33 out of
100 configurations ($\sim 33\%$) have a $d$ larger than 3.5 lattice spacings.
For the smaller lattice this is 27 out of 114 configurations ($\sim 24\%$).
Thus for both lattice sizes we find that a sizable portion of the configurations
show a different location of the zero modes for periodic and anti-periodic
boundary conditions. This answers the question which we posed at the
end of our introduction to the positive, and we have established, that also 
SU(2) shows hopping zero modes when the boundary conditions are switched.

\end{document}